\documentclass[conference]{IEEEtran}
\IEEEoverridecommandlockouts
\usepackage{cite}
\usepackage{amsmath,amssymb,amsfonts}
\usepackage{algorithmic}
\usepackage{graphicx}
\usepackage{textcomp}
\usepackage{xcolor}

\usepackage{verbatim}
\usepackage{amsmath}
\usepackage{amssymb}
\usepackage{algorithm}
\usepackage{stfloats}
\usepackage{subfigure}
\usepackage{booktabs}

\def\BibTeX{{\rm B\kern-.05em{\sc i\kern-.025em b}\kern-.08em
    T\kern-.1667em\lower.7ex\hbox{E}\kern-.125emX}}
\begin{document}

\title{Linear MIMO Precoders Design for Finite Alphabet Inputs via Model-Free Training\\
}


\author{\IEEEauthorblockN{Chen Cao, Biqian Feng, Yongpeng Wu, Derrick Wing Kwan Ng, and Wenjun Zhang}
	\thanks{C. Cao, B. Feng, Y. Wu and W. Zhang are with the \mbox{Department} of Electronic Engineering, Shanghai Jiao Tong University, Shanghai 200240, China (e-mail: cici$\_$sjtu@sjtu.edu.cn; fengbiqian@sjtu.edu.cn; \mbox{yongpeng.wu@sjtu.edu.cn}; zhangwenjun@sjtu.edu.cn.) (Corresponding author: Yongpeng Wu.)
		
	D. W. K. Ng is with the School of Electrical Engineering and Telecommunications, University of New South Wales, Sydney, NSW 2052, Australia (e-mail: w.k.ng@unsw.edu.au).
	
}
}

\maketitle

\begin{abstract}
This paper investigates a novel method for designing linear precoders with finite alphabet inputs based on autoencoders (AE) without the knowledge of the channel model. By model-free training of the autoencoder in a multiple-input multiple-output (MIMO) system, the proposed method can effectively solve the optimization problem to design the precoders that maximize the mutual information between the channel inputs and outputs, when only the input-output information of the channel can be observed. Specifically, the proposed method regards the receiver and the precoder as two independent parameterized functions in the AE and alternately trains them using the exact and approximated gradient, respectively. Compared with previous precoders design methods, it alleviates the limitation of requiring the explicit channel model to be known. Simulation results show that the proposed method works as well as those methods under known channel models in terms of maximizing the mutual information and reducing the bit error rate.

\end{abstract}


\section{Introduction}
\IEEEPARstart{L}{inear} precoding prevails as a key technology in multiple-input multiple-output (MIMO) systems, which can improve both the transmission rate and communication quality \cite{jing1}. To achieve the MIMO system capacity, various precoding methods based on the Gaussian channel input assumption have been proposed. Among them, water-filling (WF) \cite{jing3} has been proved to be the theoretically optimal precoding method, which can achieve the channel capacity. However, in practical systems, channel inputs are usually drawn from finite alphabets, such as phase-shift keying (PSK) signals, quadrature-amplitude modulation (QAM) signals, etc. Therefore, those methods based on the Gaussian input assumption would inevitably cause performance loss, making it more valuable to study the design methods of precoders taking into account the impacts of finite alphabets inputs \cite{xiao20}. Under this premise, the precoding design goal is usually to maximize the channel input-output mutual information \cite{xiao20}. For instance, \cite{xiao20} proposes mercury/water-filling (MWF), which is an optimal power allocation method for independent parallel additive white Gaussian noise (AWGN) channel. Also, \cite{xiao23} have proved that the mutual information is a concave function with respect to (w.r.t.) the squared singular values of the precoding matrix if its right singular vectors are fixed. Thus, for a general vector Gaussian channel, the design of precoding matrix can be reduced to the power distribution matrix design and the right singular matrix design by the singular value decomposition (SVD) of the precoding matrix \cite{jing14}. 

In recent years, due to the widespread use of machine learning (ML), the transmitter and receiver in a communication system based on autoencoders (AE) can be optimized in pairs, rather than the separated approach as in traditional methods, which helps achieve better performance over the whole system \cite{free2}. 
Particularly, \cite{jing} has proved that the process of AE training on MIMO precoders and receivers can maximize the mutual information, by properly selecting the activation function and loss function. It is worth mentioning that the time complexity in the optimization process can be reduced significantly \cite{jing}, as the network training process avoids the explicit calculation of the mutual information and its gradient.  

Despite the promised potential performance gain, these aforementioned precoding design methods are based on the assumption that a complete channel model can be obtained at the transmitter. In fact, on the one hand, the channel model in practical systems is hard to derive or estimate and only the input and output data can be observed; on the other hand, channel estimation with errors would inevitably occur and lead to performance loss\cite{free4}. As such, \cite{free} proposes an alternating model-free algorithm for training AEs in point-to-point single-input single-output (SISO) communication systems without channel models. 
Unfortunately, \cite{free} focuses only on the AWGN and Rayleigh block-fading channels without considering the role of precoding. In addition, it does not make the channel inputs be drawn from finite alphabets. Instead, it sends the message directly to the channel with a normalized energy through the neural networks (NN). Therefore, this method is difficult to be applied in practical MIMO systems.  

In this paper, a novel AE-based design method for linear MIMO precoders with finite alphabet inputs is presented, which can be applied when the specific channel model is unknown. In such cases,  the channels are non-differentiable and thus conventional gradient-based training through backpropagation is not applicable, e.g., \cite{jing}. To circumvent this problem, we provide an iterative training algorithm to optimize the precoder and the receiver, which can be regarded as two independent parameterized functions. This model-free algorithm iterates between training receivers with the true gradients and training precoders with the estimated gradients by treating the channel inputs as random variables.

To summarize, the main contributions of this paper are as follows: i) By alleviating the limitation of existing precoding design schemes in requiring a complete channel model,  precoders can be designed under directly observable channel information, which avoids the bad influence of channel modeling error on the precoders design; ii) Jointly optimizing the receiver and the precoder in MIMO systems for the practical channel information. Simulation results show that the proposed method can achieve similar performance to the model-based algorithm in maximizing the mutual information and has admitted a good performance in reducing the bit error rate (BER).  


\emph{Notations:} Boldface uppercase (lowercase) letters denote matrices (column vectors). $\mathbf{I}_{n}$ denotes the $n\times n$ identity matrix. $\mathbb{R}(\mathbb{C})$ is the set of real (complex) numbers. $\mathrm{Re}(\cdot)$ and $\mathrm{Im}(\cdot)$ denote the real part and the imaginary part of the complex matrix, respectively. $\mathcal{N}(m,s)$ is the Gaussian distribution with mean $m$ and covariance $s$. $\mathcal{H}(\cdot)$ denotes the information entropy of the random variables.  $\mathbb{E}_{\mathbf{x}}(\cdot)$ represents the expectation w.r.t. $\mathbf{x}$. $\rm{Tr}(\cdot)$ denotes the trace of the matrix. $\rm{Bdiag}(\cdot)$ means a block diagonal matrix. The gradient and Jacobian operators w.r.t. the set of parameters $\boldsymbol{\theta}$ are both denoted by $\nabla_{\boldsymbol{\theta}}$; and the superscripts $(\cdot)^{\mathsf{T}}$, $(\cdot)^{\mathsf{H}}$, and $\Vert\cdot\Vert_2$ represent transpose, conjugate transpose operations, and $l_2$-norm, respectively. 
 
\section{System Model}

Considering a MIMO communication system which has $N_{t}$ antennas at the transmitter and $N_{r}$ antennas at the receiver, the received signal can be expressed as:
\begin{equation}
	\setlength{\abovedisplayskip}{3pt}
	\setlength{\belowdisplayskip}{3pt}
	\mathbf{y} = \mathbf{HGx} + \mathbf{n}\label{eq},
\end{equation}
where $\mathbf{H}\in\mathbb{C}^{N_r\times N_t}$ is the channel matrix, $\mathbf{G}\in\mathbb{C}^{N_t\times N_t}$ is the linear precoder, $\mathbf{n}\in\mathbb{C}^{N_r\times1}$ is the circularly symmetric white Gaussian noise whose covariance matrix is $\sigma^2\mathbf{I}_{N_r}$, and $\mathbf{x}\in\mathbb{C}^{N_t\times1}$ is the input signal of zero mean and covariance $\mathbb{E}\{\mathbf{x}\mathbf{x}^{\mathsf{H}}\}=\mathbf{I}_{N_t}$. Based on the assumption of finite alphabet inputs, the input signal $\mathbf{x}$ is equiprobably drawn from a set $\mathcal{M}$ of discrete constellations such as $M$-ary PSK or QAM with $|\mathcal{M}| = M^{N_t}$.

Aiming at maximizing the mutual information between channel input $\mathbf{x}$ and output $\mathbf{y}$, the design problem of a precoder with the finite alphabet inputs can be formulated as: 
\begin{equation}
	\setlength{\abovedisplayskip}{3pt}
	\setlength{\belowdisplayskip}{3pt}
	\begin{split}
		&\underset{\mathbf{G}}{\mathrm{maximize}}\quad\mathcal{I}(\mathbf{x};\mathbf{y}) \\	
		&\quad{\rm s.t.}\quad{\rm Tr}\{\mathbf{G}^{\mathsf{H}}\mathbf{G}\}\leq N_t,
	\end{split}
	\label{p1}
\end{equation}
where the mutual information can be described as \cite{jing5}:
\begin{equation}
	\setlength{\abovedisplayskip}{3pt}
	\setlength{\belowdisplayskip}{3pt}
	\mathcal{I}(\mathbf{x};\mathbf{y}) = N_t\log_2M - \frac{1}{M^{N_t}}\sum_{m  =1}^{M^{N_t}}\mathbb{E}_{\mathbf{n}}\left\{\log_2\sum_{k=1}^{M^{N^t}}e^{-d_{m,k}}\right\},
\end{equation}
with $d_{m,k} = \sigma^{-2}(\|\mathbf{HG}(\mathbf{x}_m-\mathbf{x}_k)+\mathbf{n}\|_2^2-\|\mathbf{n}\|_2^2 ).$

Nevertheless, it is difficult to obtain the channel model $\mathbf{H}$ \cite{ZF}, and an inaccurate channel model may cause performance loss \cite{free4}.

\section{Algorithm Design}
\begin{figure}[h]
	\centering{\includegraphics[scale = 0.35]{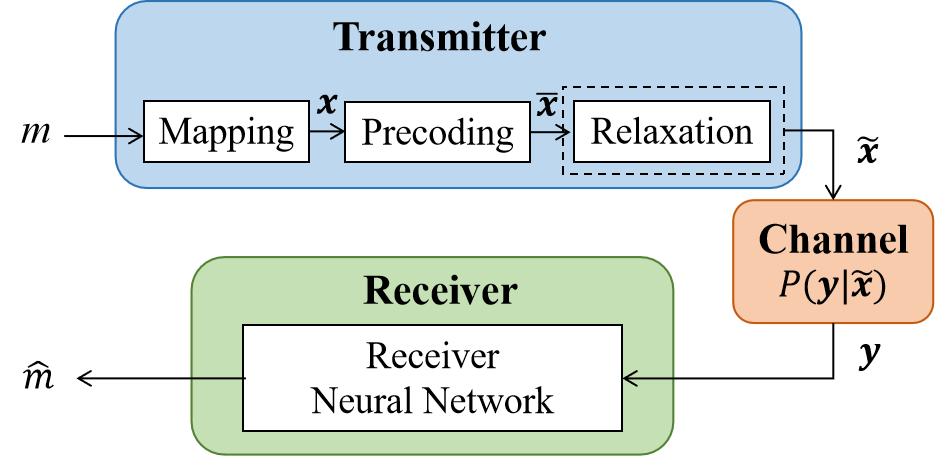}}
	\caption{A MIMO communication system based on an AE in model-free training. The dotted box indicates that the module of relaxation is only existed when training precoders.}
	\label{fig1}
\end{figure}

The MIMO communication system based on the AE presented in this paper includes three parts: the transmitter, the channel, and the receiver, as shown in \mbox{Fig. \ref{fig1}}. First, the transmitter part is equivalent to the signal modulation and precoding process. Second, the channel corresponds to a random system \cite{free}, whose output $\mathbf{y}$ follows $p(\mathbf{y}|\mathbf{x})$, i.e., a conditional probability distribution w.r.t. input $\mathbf{x}$. In the AE, it can be regarded as a layer of untrainable network parameters, which transmits signals to the receiver through forward propagation. Finally, the receiver recovers the transmitted information from the received signal through NNs. 
In this paper, both the precoding and receiver network are the parts to be optimized by minimizing the loss function.  

\subsection{Model-free Training Process Overview of Precoders Design based on AE}

In the proposed method, we regard the precoder and the receiver as two independent parameterized functions to optimize: (i) the precoder is presented by the function $f_{\boldsymbol{\theta}_{P}}^{(P)}(\mathbf{x})=\boldsymbol{\theta}_{P}\mathbf{x}$, where $\boldsymbol{\theta}_P =  \begin{bmatrix}
	\mathrm{Re}(\mathbf{G})&-\mathrm{Im}(\mathbf{G})\\
	\mathrm{Im}(\mathbf{G})&\mathrm{Re}(\mathbf{G})
\end{bmatrix}$ 
is the parameter matrix of the precoder; (ii) the receiver is implemented as $f_{\boldsymbol{\theta}_{R}}^{(R)} : \mathbb{C}^{N_t} \rightarrow \{\mathbf{p}\in\mathbb{R}_+^{|\mathcal{M}|} |\sum_{i}^{|\mathcal{M}|}p_i = 1 \}$, where $\boldsymbol{\theta}_R$ is the parameter vector of receivers, and $\mathbf{p}$ is the probability vector of the transmitted information. Since the NN implementation is limited to the range of real numbers, in this paper, the real and imaginary parts of complex signals involved need to be reshaped into real number vectors before further processing. 

Considering the channel follows a conditional probability distribution w.r.t. the channel input, the loss function of the system can be expressed as:
\begin{equation}
	\setlength{\abovedisplayskip}{3pt}
	\setlength{\belowdisplayskip}{3pt}
	\begin{split}
		\mathcal{L}(\boldsymbol{\theta}_P,\boldsymbol{\theta}_R) =\; &  \mathbb{E}_{\mathbf{x}}\left\{\int l(f_{\boldsymbol{\theta}_{R}}^{(R)}(\mathbf{y}),\mathbf{x})p(\mathbf{y}|\bar{\mathbf{x}})d\mathbf{y}\right\}
		\\
		\approx\;&\frac{1}{S}\sum_{i=1}^{S}l(f_{\boldsymbol{\theta}_{R}}^{(R)}(\mathbf{y}^{(i)}),\mathbf{x}^{(i)}),
	\end{split}
\end{equation}
where $\bar{\mathbf{x}}= f_{\boldsymbol{\theta}_{P}}^{(P)}(\mathbf{x})$ represents the precoded signal, $\mathbf{y} \in \mathbb{R}^{2N_t\times1}$ denotes the received signal, as well as the channel output, and $l(f_{\boldsymbol{\theta}_{R}}^{(R)}(\mathbf{y}^{(i)}),\mathbf{x}^{(i)})$ is the sample loss function defined as the categorical cross-entropy (CE) \cite{free,jing} between the $i$th input signal $\mathbf{x}^{(i)}$ and the $i$th received signal $\mathbf{y}^{(i)}$, $S$ is the batch size of training samples. The approximation in (4) means that we can use the sample mean to estimate the mathematical expectation.

Then, the corresponding optimization problem is formulated as:
\begin{equation}
	\setlength{\abovedisplayskip}{3pt}
	\setlength{\belowdisplayskip}{3pt}
	\begin{split}
		&\underset{\boldsymbol{\theta}_{P},\boldsymbol{\theta}_{R}}{\rm{\mathrm{minimize}}}\quad\mathcal{L}(\boldsymbol{\theta}_{P},\boldsymbol{\theta}_{R})\\	
		&{\rm s.t.}\quad{\rm Tr}\left\{\mathbf{G}^{\mathsf{H}}\mathbf{G}\right\}\leq N_t.
	\end{split}
	\label{p5}
\end{equation}
When Softmax is adopted as the activation function of the output layer, inspired by \cite{jing}, then we have 
\begin{equation}
	\setlength{\abovedisplayskip}{3pt}
	\setlength{\belowdisplayskip}{3pt}
	\begin{split}		\mathcal{L}(\boldsymbol{\theta}_{P},\boldsymbol{\theta}_{R}) &= \mathbb{E}_{\mathbf{y}}\left\{-\sum^{|\mathcal{M}|}_{m = 1}p(\mathbf{x}=\mathbf{x}_m|\mathbf{y})\log_2(f_{\boldsymbol{\theta}_{R}}^{(R)}(\mathbf{y})_m)\right\}
		\\ 
		&\geq \mathcal{H}(\mathbf{x}|\mathbf{y}).
	\end{split}
\end{equation}
Since $\mathcal{I}(\mathbf{x}|\mathbf{y}) = \mathcal{H}(\mathbf{x}) - \mathcal{H}(\mathbf{x}|\mathbf{y})$, the problem in (\ref{p5}) is nearly the same as the problem in (\ref{p1}), i.e., the precoder obtained by training the AE can also maximize the mutual information.

The training process requires the derivative of the loss function, i.e.,$\nabla_{(\boldsymbol{\theta}_R,\boldsymbol{\theta}_{P})}\mathcal{L} = [(\nabla_{\boldsymbol{\theta}_R}\mathcal{L})^{\mathsf{T}},(\nabla_{\boldsymbol{\theta}_P}\mathcal{L})^{\mathsf{T}}]^{\mathsf{T}}$. To enable the training for the proposed framework in the absence of the channel model, inspired by \cite{free}, we present an alternating training method with the exact gradient and the approximated gradient, respectively, as shown in Algorithm \ref{alg:out_iter}.
It should be noted that since the loss function in this algorithm is defined as a categorical CE function, the transmitted information should be in the form of one-hot and then mapped to corresponding discrete constellation points, as shown in \mbox{Fig. \ref{fig1}}. We adopt the Adam optimizer \cite{jing29} for training.

\begin{algorithm}[t]
	\caption{Autoencoder Training}
	\label{alg:out_iter}
	\begin{algorithmic}[1]
		\STATE \textbf{Initialization:} Set $\boldsymbol{\theta}_{P}$ and $\boldsymbol{\theta}_{R}$ to some random values, such that ${\rm Tr}\{\mathbf{G}^H\mathbf{G}\}\leq N_t$.
		\WHILE{stop criterion not met}
		\STATE Train the Receiver;
		\STATE Train the Precoder;
		\ENDWHILE
	\end{algorithmic}
\end{algorithm}
\addtolength{\topmargin}{0.01in}
\subsection{Receiver Training}
According to (4), the gradient of $\mathcal{L}$ w.r.t. $\boldsymbol{\theta}_{R}$ can be expressed as:
\begin{equation}
	\setlength{\abovedisplayskip}{3pt}
	\setlength{\belowdisplayskip}{3pt}
	\begin{split}
	\nabla_{\boldsymbol{\theta}_R}\mathcal{L} &= \mathbb{E}_{\mathbf{x},\mathbf{y}}\left\{\nabla_{\boldsymbol{\theta}_R}l(f_{\boldsymbol{\theta}_{R}}^{(R)}(\mathbf{y}),\mathbf{x})\right\}
	\\
	&\approx \frac{1}{S}\sum_{i=1}^{S}\nabla_{\boldsymbol{\theta}_R}l(f_{\boldsymbol{\theta}_{R}}^{(R)}(\mathbf{y}^{(i)}),\mathbf{x}^{(i)}).
	\end{split}
\end{equation}
It can be seen that there is no need to acquire the channel model, as the calculation process only needs to sample the received signal.

\begin{algorithm}[t]
	\caption{Receiver Training}
	\label{alg:rx_iter}
	\begin{algorithmic}[1]
		\FOR{$i=1$ \textbf{to} $\mathrm{iter}_{\max}$}
		\STATE Generate $S$ independent input information $\mathbf{M}$;
		\STATE Map $\mathbf{M}$ to finite alphabet signal $\mathbf{X}$;
		\STATE $\bar{\mathbf{X}} \leftarrow f_{\boldsymbol{\theta}_{P}}^{(P)}(\mathbf{X})$;
		\STATE Send $\bar{\mathbf{X}}$ to channel and get $\mathbf{Y}$;
		\STATE $\hat{\mathbf{M}} \leftarrow f_{\boldsymbol{\theta}_{R}}^{(R)}(\mathbf{Y})$;
		\STATE Compute loss $\mathcal{L}$ and $\nabla_{\boldsymbol{\theta}_R}\mathcal{L}$
		\STATE Update $\boldsymbol{\theta}_{R}$ with one Adam step;
		\ENDFOR
		\STATE \textbf{Output:} $\boldsymbol{\theta}_{R}$
	\end{algorithmic}
\end{algorithm}
Algorithm \ref{alg:rx_iter} is the detailed algorithm for training the receiver. First, the signal source randomly generates a batch of one-hot transmitted information $\mathbf{M}$, which is a $|\mathcal{M}|$-by-$S$ matrix and then mapped to the finite alphabet input signals $\mathbf{X}$. Next, the signals are multiplied by $\boldsymbol{\theta}_{P}$ to obtain the precoded signals $\bar{\mathbf{X}}$, and then $\bar{\mathbf{X}}$ enters the channel for transmission. The receiver acquires the channel output $\mathbf{Y}$. After $\mathbf{Y}$ is fed to the receiver network, a batch of probability matrix $\hat{\mathbf{M}} \in \mathbb{R}^{|\mathcal{M}|\times S}$ over the transmitted information are given for the calculating loss $\mathcal{L}$. Finally, $\boldsymbol{\theta}_{R}$ can perform a one-step update using $\nabla_{\boldsymbol{\theta}_R}\mathcal{L}$ by (7).
As the channel model is unknown, lines 4, 5 in Algorithm \ref{alg:rx_iter} only carry out forward propagation and do not need to record the gradient. 
\subsection{Precoder Training}
According to (4), the gradient of $\mathcal{L}$ w.r.t. $\boldsymbol{\theta}_{P}$ is:

\begin{equation}
	\setlength{\abovedisplayskip}{3pt}
	\setlength{\belowdisplayskip}{3pt}
	\begin{split}
		\nabla_{\boldsymbol{\theta}_P}\mathcal{L} = \mathbb{E}_{\mathbf{x}}\bigg\{\int l(f_{\boldsymbol{\theta}_{R}}^{(R)}(\mathbf{y}),\mathbf{x})	\nabla_{\bar{\mathbf{x}}}p(\mathbf{y}|\bar{\mathbf{x}})\nabla_{\boldsymbol{\theta}_P}f_{\boldsymbol{\theta}_{P}}^{(P)}(\mathbf{y})d\mathbf{y}\bigg\}.
		\end{split}
\end{equation}
Since the channel model $p(\mathbf{y}|\bar{\mathbf{x}})$ is unknown and $\nabla_{\bar{\mathbf{x}}}p(\mathbf{y}|\bar{\mathbf{x}})$ cannot be calculated, inspired by \cite{free}, we resort to another approach that relax the channel input $\bar{\mathbf{x}}$ into a random variable $\tilde{\mathbf{x}}$, which follows a distribution of $\pi_{\bar{\mathbf{x}}} = \delta(\tilde{\mathbf{x}} - \bar{\mathbf{x}})$, where $\delta$ refers to the delta distribution. The position of the relaxation operation in the algorithm process is shown in the dotted box in \mbox{Fig.  \ref{fig1}}. Then, the relaxed system loss can be expressed as:
\begin{equation}
	\setlength{\abovedisplayskip}{3pt}
	\setlength{\belowdisplayskip}{3pt}
	\mathcal{\hat{L}} = \mathbb{E}_{\mathbf{x}}\left\{\iint l(f_{\boldsymbol{\theta}_{R}}^{(R)}(\mathbf{y}),\mathbf{x})p(\mathbf{y}|\tilde{\mathbf{x}})\pi_{\bar{\mathbf{x}}}(\tilde{\mathbf{x}})d\tilde{\mathbf{x}}d\mathbf{y}\right\}.
\end{equation}
Besides, $\nabla_{\boldsymbol{\theta}_P}\mathcal{L}$ can be approximated by the following expression:
\begin{equation}
	\setlength{\abovedisplayskip}{3pt}
	\setlength{\belowdisplayskip}{3pt}
	\begin{split}
		\nabla_{\boldsymbol{\theta}_P}\mathcal{\hat{L}} &= \mathbb{E}_{\mathbf{x}}\left\{\iint l(f_{\boldsymbol{\theta}_{R}}^{(R)}(\mathbf{y}),\mathbf{x})p(\mathbf{y}|\tilde{\mathbf{x}})\nabla_{\boldsymbol{\theta}_P}\pi_{\bar{\mathbf{x}}}(\tilde{\mathbf{x}})d\tilde{\mathbf{x}}d\mathbf{y}\right\}		
		\\
		&=\mathbb{E}_{\mathbf{x},\tilde{\mathbf{x}},\mathbf{y}}\left\{l(f_{\boldsymbol{\theta}_{R}}^{(R)}(\mathbf{y}),\mathbf{x})\nabla_{\boldsymbol{\theta}_P}f_{\boldsymbol{\theta}_{P}}^{(P)}(\mathbf{y})\nabla_{\boldsymbol{\bar{\mathbf{x}}}}\log\pi_{\bar{\mathbf{x}}}(\tilde{\mathbf{x}})\right\}	
		\\
		&\approx \frac{1}{S}\sum_{i=1}^{S}l(f_{\boldsymbol{\theta}_{R}}^{(R)}(\mathbf{y}^{(i)}),\mathbf{x}^{(i)})\cdot\\
		&\quad\ \ \nabla_{\boldsymbol{\theta}_P}\log(\hat{\pi}_{\bar{\mathbf{x}},\sigma_{\pi}}(\tilde{\mathbf{x}}^{(i)}))|_{{\bar{\mathbf{x}}} = f_{\boldsymbol{\theta}_P}^{(P)}(\mathbf{x}^{(i)})} ,
	\end{split}
\end{equation}
where $\hat{\pi}_{\bar{\mathbf{x}},\sigma_{\pi}}$ represents the approximated probability distribution function of $\pi_{\bar{\mathbf{x}}}$, whose variance is $\sigma_{\pi}^2$, since $\pi_{\bar{\mathbf{x}}}$ is non-differentiable. As such, the function requiring the derivative of $\boldsymbol{\theta}_P$ is subtly transformed from the unknown channel model $p(\mathbf{y}|\mathbf{x})$ to the known distribution $\hat{\pi}$. Through the estimated gradient in (10), we can complete the training process of precoders without the need of the channel model. 

\begin{algorithm}[t]
	\caption{Precoder Training}
	\label{alg:tx_iter}
	\begin{algorithmic}[1]
		\FOR{$i=1$ \textbf{to} $\mathrm{iter}_{\max}$}
		\STATE Generate $S$ independent input information $\mathbf{M}$;
		\STATE Map $\mathbf{M}$ to finite alphabet signal $\mathbf{X}$;
		\STATE $\bar{\mathbf{X}} \leftarrow f_{\boldsymbol{\theta}_{P}}^{(P)}(\mathbf{X})$;
		\STATE Relax $\bar{\mathbf{X}}$ to random variable $\tilde{\mathbf{X}}$
		\STATE Send $\tilde{\mathbf{X}}$ to channel and acquire $\mathbf{Y}$;
		\STATE $\hat{\mathbf{M}} \leftarrow 		f_{\boldsymbol{\theta}_{R}}^{(R)}(\mathbf{Y})$;
		\STATE Update $\boldsymbol{\theta}_{P}$ with one Adam step 
		\STATE Scale $\boldsymbol{\theta}_{P}$, such that ${\rm Tr}\{\mathbf{G}^H\mathbf{G}\}\leq N_t$;
		\STATE Compute per-sample loss $l$ and   $\nabla_{\boldsymbol{\theta}_P}\mathcal{\tilde{L}}$
		\ENDFOR
		\STATE \textbf{Output:} $\boldsymbol{\theta}_{P}$
	\end{algorithmic}
\end{algorithm}
\addtolength{\topmargin}{0.01in}
Algorithm \ref{alg:tx_iter} is the detailed algorithm for training the precoder. The signal source randomly generates  the transmitted information $\mathbf{M}$ and maps them to the input signals $\mathbf{X}$. After precoding, $\bar{\mathbf{X}}$ can be obtained. For the approximated gradient calculation, it is necessary to relax $\bar{\mathbf{X}}$ into random variables $\tilde{\mathbf{X}}$, and then sends the random variables to the channel for transmission to obtain the received signals $\bar{\mathbf{Y}}$. The receiver then establishes the probability matrix of the received signals over the transmitted information and then calculates the per-sample loss $l$. However, at this time, $l$ does not directly calculate the gradient of $\boldsymbol{\theta}_{P}$, as shown in (10), and $\hat{\pi}_{\bar{\mathbf{x}},\sigma_{\pi}}$ should be used to calculate the approximated gradient. Then, $\boldsymbol{\theta}_{P}$ can perform a step update through 	$\nabla_{\boldsymbol{\theta}_P}\mathcal{\hat{L}}$, which still needs to satisfy the power constraints of the precoders. 
In order to ensure that $\nabla_{\boldsymbol{\theta}_P}\mathcal{\hat{L}}$ can be estimated and receiver parameters are not affected, lines 6, 7 in Algorithm \ref{alg:tx_iter} only carry out forward propagation and do not need to record the gradient. In other words, the channel in this algorithm is only used to observe its input and the corresponding output, and the complete channel model itself does not participate in the precoders design process.  

\begin{figure}
	\centering\subfigure[]{
		\label{fig:subfig:a} 
		\includegraphics[width = 1.5in]{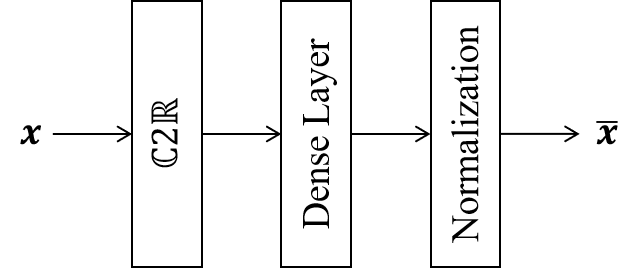}}
	\subfigure[]{
		\label{fig:subfig:b} 
		\includegraphics[width = 2in]{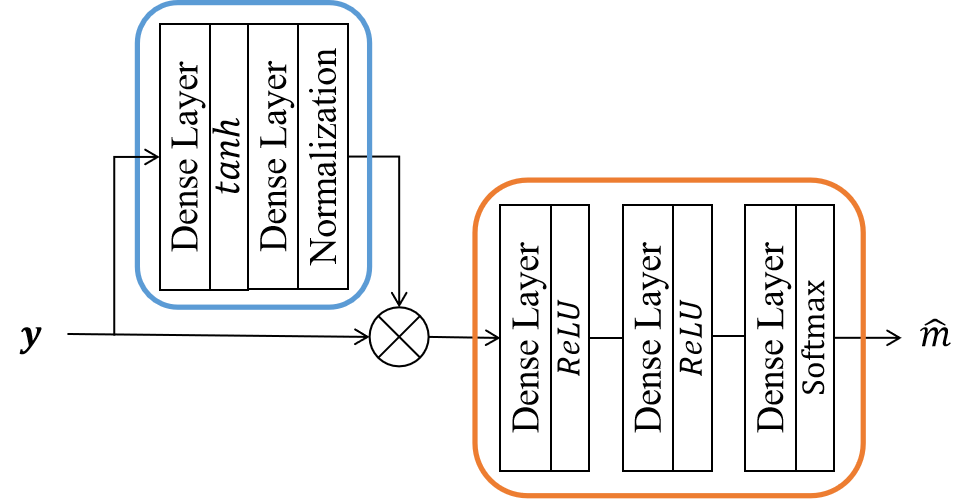}}
	\caption{Network structure: (a) Precoder structure; (b) Receiver structure,  where the blue and orange box indicate the ``equalization" network and decision network, respectively. }
	\label{fig:subfig}
	\label{fig2}

\end{figure}

\subsection{Network Structure}

\mbox{Fig. \ref{fig2}} shows the network structure of the algorithm proposed in this paper, where \mbox{Fig. \ref{fig:subfig:a}} is the precoding part and \mbox{Fig. \ref{fig:subfig:b}} is the receiver part. As shown in the figure, the precoding matrix $\mathbf{G}$ is implemented by a linear layer without bias, whose parameters are $\boldsymbol{\theta}_P$. The network structure of the receiver is mainly composed of two parts: the first one is the ``equalization" network, which consists of two linear layers, and the activation function between is the $tanh$ function; the second one is the decision network, which is composed of multiple linear layers, between which the activation function is ReLU, and Softmax at the output layer. It should be noted that since the proposed method is used in the case of unknown channel models, the receiver may show poor performance \cite{free} without prior information. Therefore, in order to improve the accuracy of the receiver, the ``equalization" network is added to extract part of implicit prior channel information $\hat{\mathbf{H}} \in \mathbb{C}^{N_t\times S}$ from the received signals. Then, it calculates the product of $\frac{1}{\|\hat{\mathbf{H}}\|_2^2} \begin{bmatrix}
	\mathrm{Re}(\hat{\mathbf{H}})&\mathrm{Im}(\hat{\mathbf{H}})\\
	-\mathrm{Im}(\hat{\mathbf{H}})&\mathrm{Re}(\hat{\mathbf{H}})
\end{bmatrix}$ and the received signal $\mathbf{y}$, similar to the channel equalization in traditional communication systems. Next, the received signals after ``equalization" can be exploited to establish the probability matrix $\hat{\mathbf M}$ over the input information through the decision network, and finally the loss function and the corresponding gradient can be calculated to train the whole network. It should be noted that $\hat{\mathbf{H}}$ is just an intermediate variable of the receiver,  which does not correspond strictly to the true channel model. Instead, it can be regarded as the implicit channel information, explaining the reason why such a structure can improve the performance of the receiver. The method itself does not need to assume the channel model in advance.    

\subsection{Improvement and Some Applications of the Proposed Linear Precoders Design Method}

As described in Section III-A, when Softmax and the categorical CE are adopted, the sizes of layers in the AE grow exponentially with the number of $N_t$, which is very expensive in large
systems.
Therefore, inspired by \cite{jing}, we use $N_t\log_2(M)$ bits to represent one training sample. Then, the loss function and the activation function at the last layer of the receiver will be adjusted to binary CE function and Sigmoid accordingly. Such a choice may make the process of training AE equivalent to maximizing the lower bound of mutual information \cite{jing}. Even so, the proposed method is easier to generalize to more complex scenarios, such as MIMO Multiple Access Channels (MAC) with finite discrete inputs, MIMO orthogonal frequency-division multiplexing (MIMO-OFDM) systems, etc.
 The accurate channel models of these systems are complex, while the assumptions of the channel models are relatively simple in the traditional research methods, or the channel model contains some non-differentiable components (such as preamble insertion in MIMO-OFDM), so the simulation results may differ greatly from the actual performance.  The proposed method without the knowledge of an accurate channel model would have inherent advantages in these systems and may alleviate this problem to some extent.  

For example, in a $K$-user MIMO MAC communication system, considering the signal model:
\begin{equation}
	\setlength{\abovedisplayskip}{3pt}
	\setlength{\belowdisplayskip}{3pt}
	\begin{split}
	\mathbf{y} = \mathbf{H}_1\mathbf{G}_1\mathbf{x}_1 + \mathbf{H}_2\mathbf{G}_2\mathbf{x}_2 + \cdots+ \mathbf{H}_K\mathbf{G}_K\mathbf{x}_K + \mathbf{n}\label{eq},	
	\end{split}
\end{equation}
where 
$\mathbf{H}_i\in\mathbb{C}^{N_r\times N_t}$ represents the complex channel matrix between the $i$th transmitter and the receiver;  
 $\mathbf{G}_i\in\mathbb{C}^{N_t\times N_t}$ is each user's precoding matrix; $\mathbf{x} = [\mathbf{x}_1^{\mathsf{T}},\mathbf{x}_2^{\mathsf{T}},\cdots,\mathbf{x}_K^{\mathsf{T}}]^{\mathsf{T}}$ contains the signal of all transmitters, assuming $\mathbf{x}_i$ of different users are independent from each other; the receiver noise $\mathbf{n}\in\mathbb{C}^{N_r\times 1}$, and $\mathbf{n} \sim \mathcal{CN}(\mathbf{0},\sigma^2\mathbf{I})$. Suppose there are $N_r$ antennas at the receiver and each user has $N_t$ transmit antennas. 

According to \cite{uplinkMU}, the boundary of the constellation-constrained capacity region can be characterized by the solution of the sum rate optimization problem.
Then, the proposed precoders design methods can be applied by: i) using $N_tK\log_2(M)$ bits to represent one training target; ii) the parameter $\boldsymbol{\theta}_{P}$ of the precoders is adjusted to Bdiag$(\boldsymbol{\theta}_{P_1},\cdots,\boldsymbol{\theta}_{P_K})$, where $\boldsymbol{\theta}_{P_i} =  \begin{bmatrix}
	\mathrm{Re}(\mathbf{G}_i)&-\mathrm{Im}(\mathbf{G}_i)\\
	\mathrm{Im}(\mathbf{G}_i)&\mathrm{Re}(\mathbf{G}_i)
\end{bmatrix}$; iii) the receiver function correspondingly becomes as $f_{\boldsymbol{\theta}_{R}}^{(R)} : \mathbb{R}^{2N_tK} \rightarrow \mathbb{R}^{N_tK\log_2(M)}$. Since the channel model of each user is unknown, the precoder and receiver parts still need the true and estimated gradients for training, respectively. It should be noted that because the optimal precoders of different users depend on each other \cite{uplinkMU}, we should iteratively optimize one user's precoder at a time with others fixed.


\begin{figure*}
	\centering
	\subfigure[Comparison under channel $\mathbf{H}_1$, BPSK]{\label{MIa}
		\includegraphics[width = 2.8in]{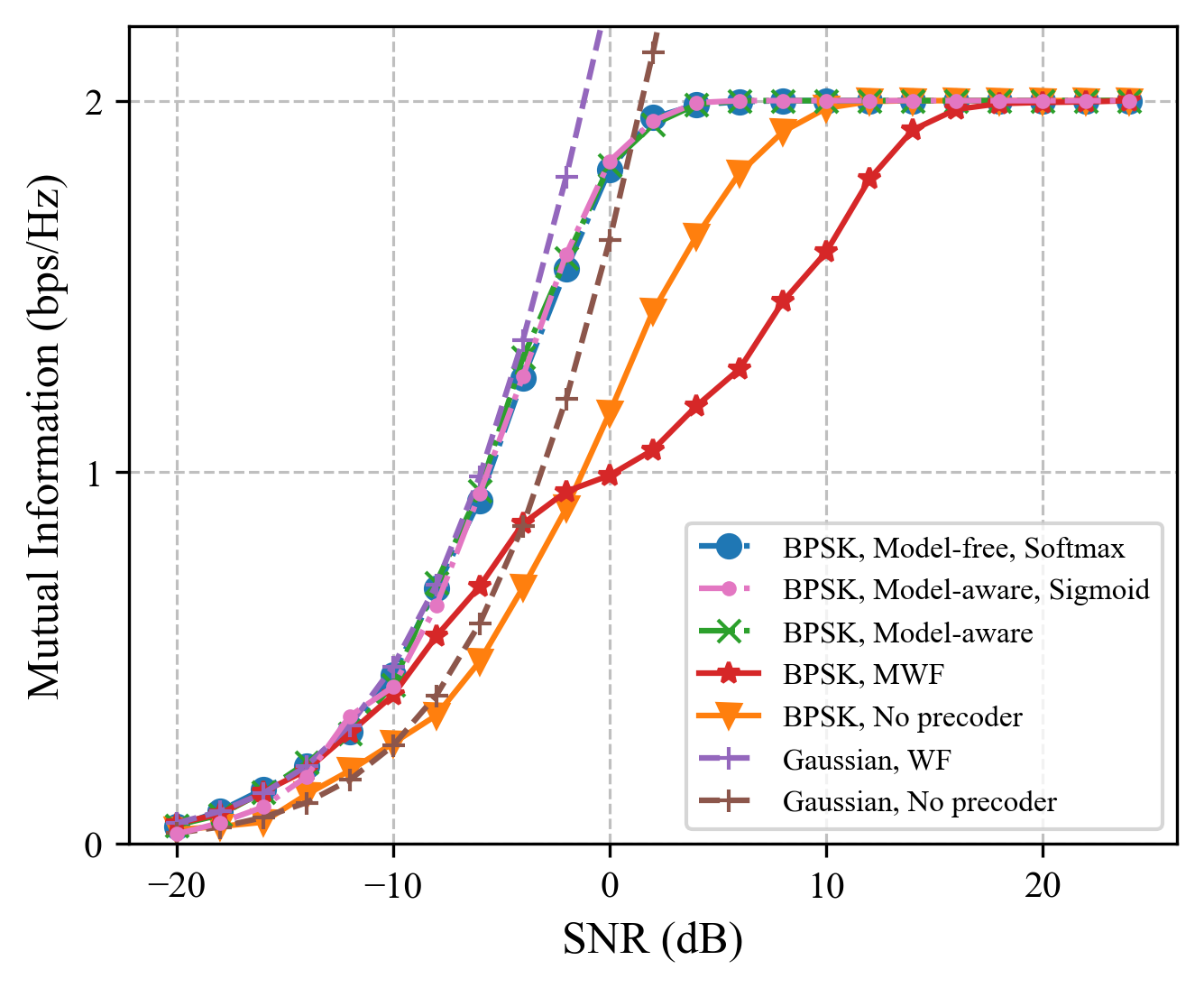}}
	\hspace{0.05\linewidth}
	\subfigure[Comparison under channel $\mathbf{H}_2$, BPSK]{\label{MIb}
		\includegraphics[width = 2.8in]{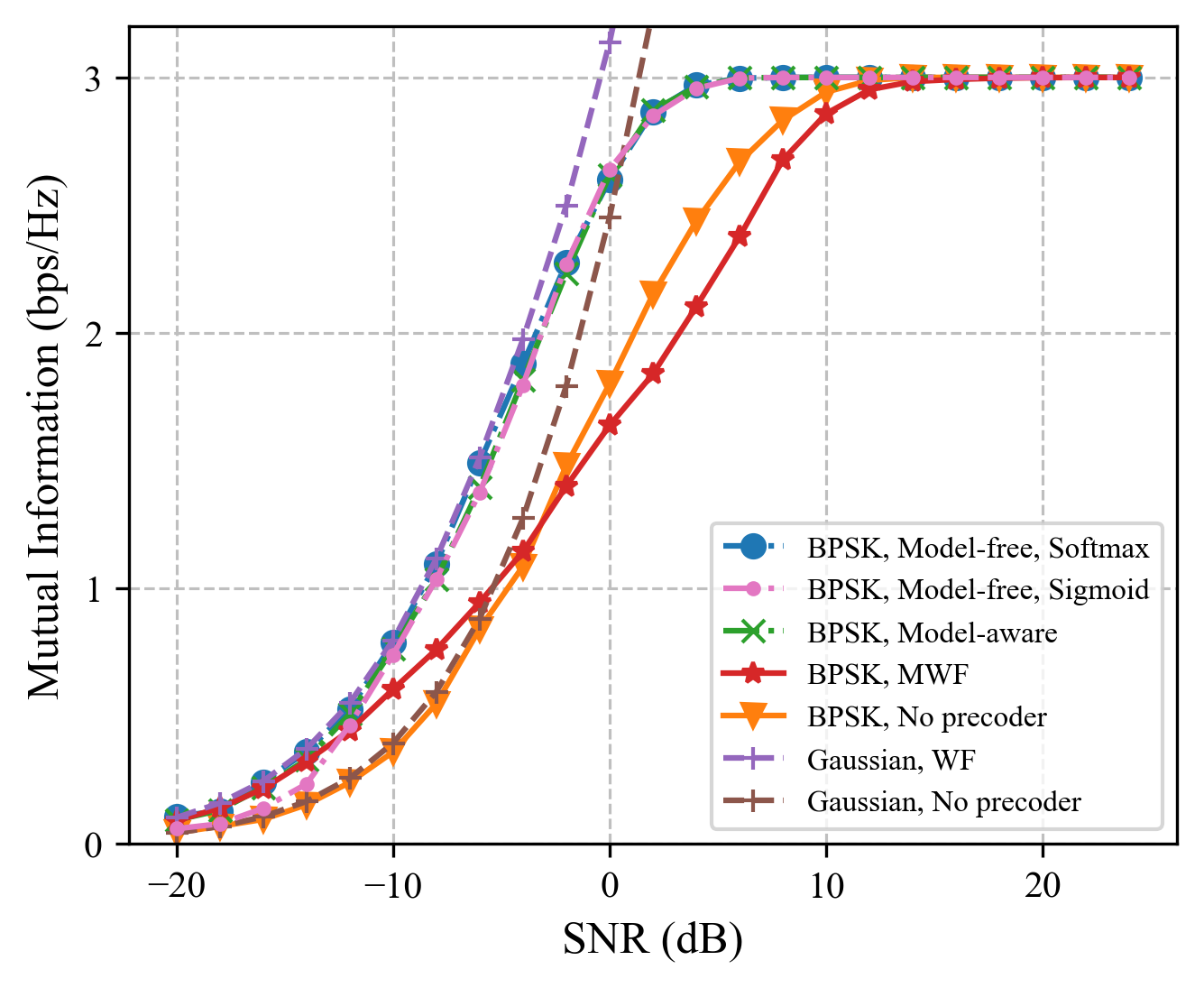}}
	\vfill
	\subfigure[Comparison under channel $\mathbf{H}_1$, QPSK]{\label{MIc}
		\includegraphics[width = 2.8in]{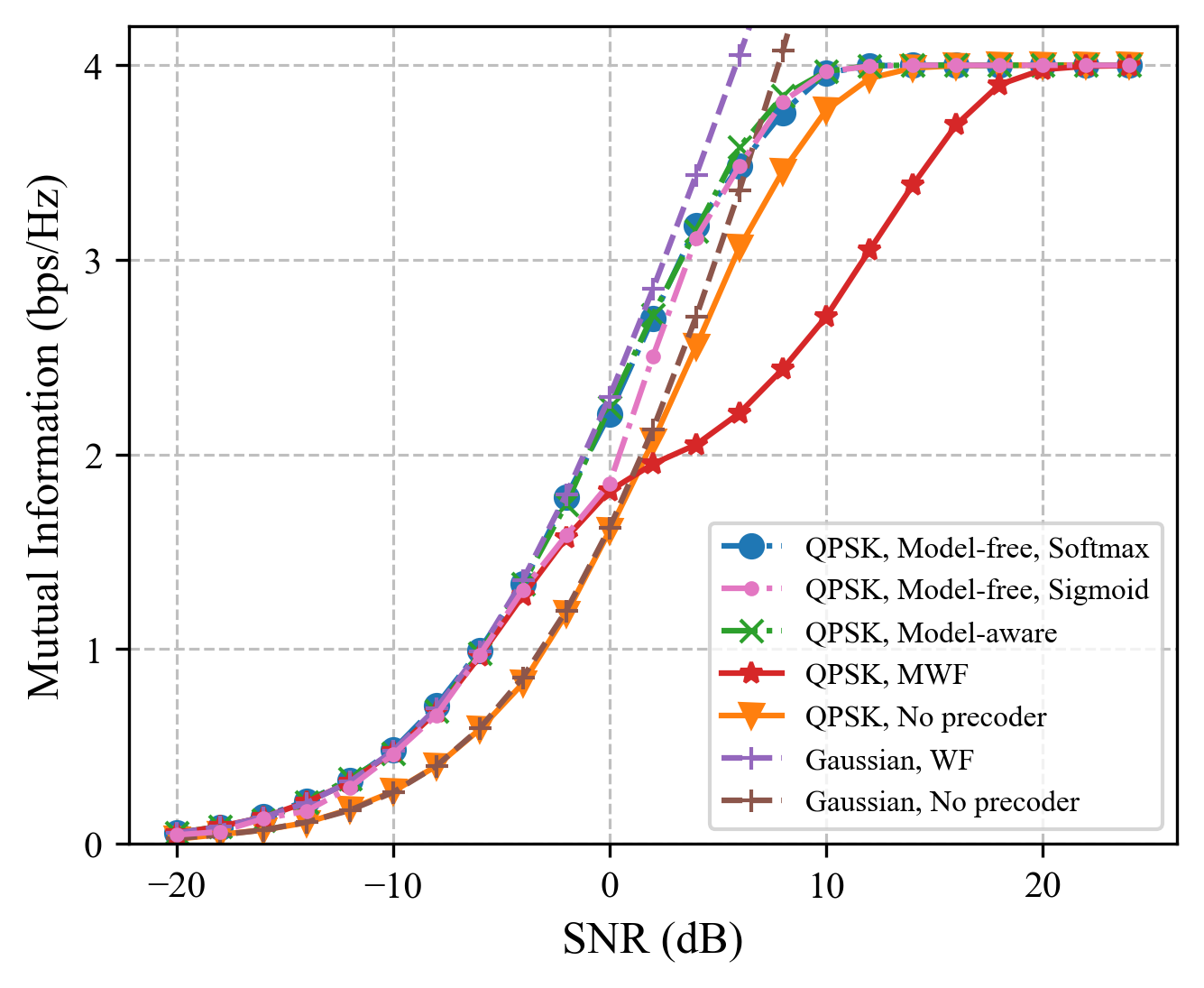}}
	\hspace{0.06\linewidth}
	\subfigure[Comparison under channel $\mathbf{H}_2$, QPSK]{\label{MId}
		\includegraphics[width = 2.8in]{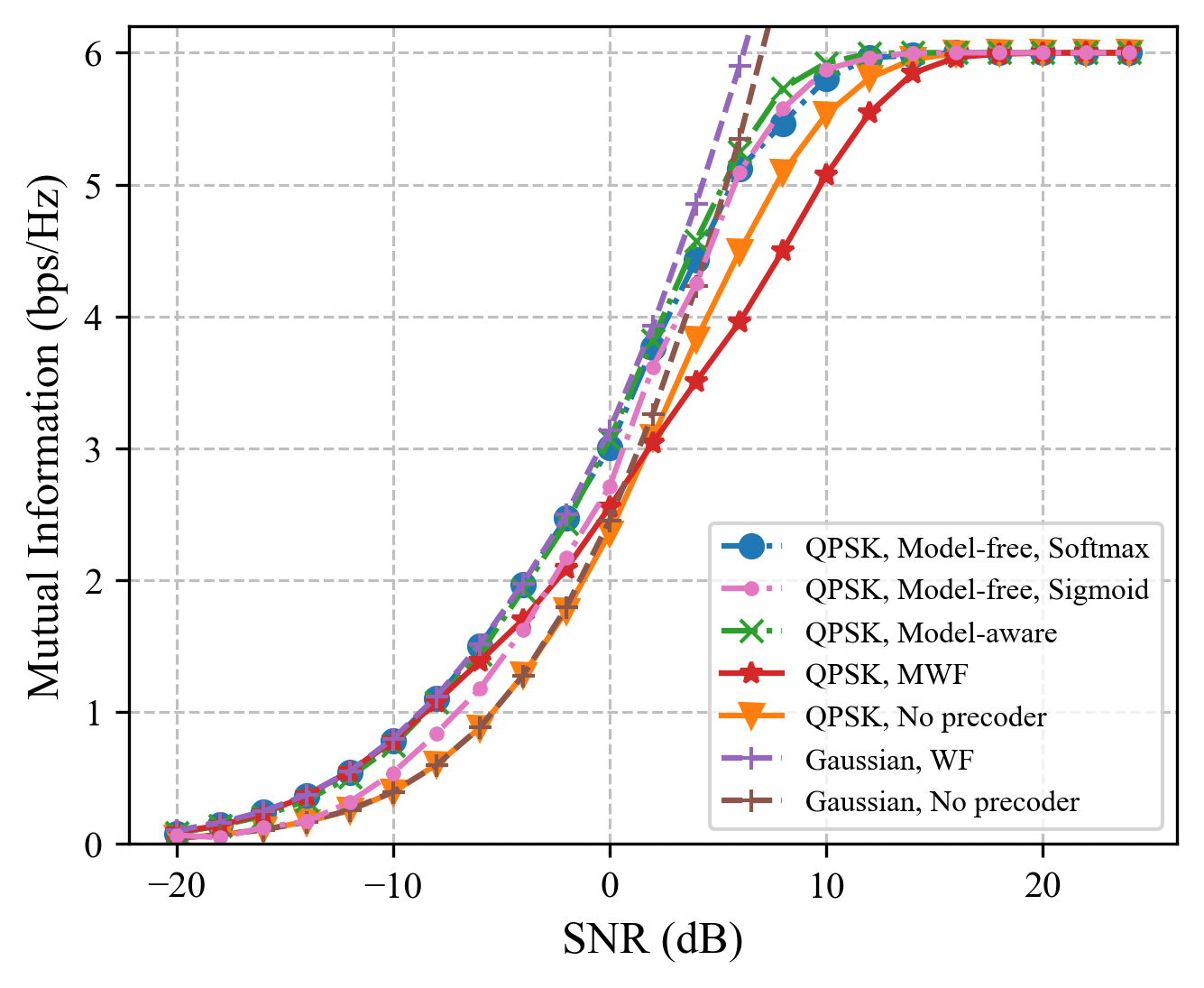}}
	\label{fig:subfig}
	\caption{Mutual information for Gaussian and finite alphabet channel inputs.}
	\label{MI}
\end{figure*}

\section{Numerical Results}

In this section, simulation results are provided to evaluate the performance of the proposed method of linear precoders design, in terms of maximizing the mutual information and the impacts on the BER of the MIMO communication system,  compared with some existing design approaches.

In the receiver part of the AE, the decision network is parameterized with 3 dense layers with size 128 each in this experiment while the ``eqalization" part has 2 dense layers with size 64 each. The maximum iteration number of the Algorithm \ref{alg:out_iter} is set as 5000 and the batchsize $S = 32$. The learning rate for Adam step on the receiver and precoder training are both set as $10^{-4}$. Also, the SNR is defined as $\frac{{\rm Tr}\{\mathbf{H}^{\mathsf{H}}\mathbf{H}\}}{N_t\sigma^2}$.

When training the precoder, we utilize the normal distribution with mean $\bar{\mathbf{x}}$ and variance $\sigma_{\pi}^2$ to relax the channel input, i.e., $ \tilde{\mathbf{x}} = \sqrt{1-\sigma_{\pi}^2}\,\bar{\mathbf{x}}+ \mathbf{w}$,    where $\mathbf{w}\sim\mathcal{N}(0,\sigma_{\pi}^2\mathbf{I})$ and the variance $\sigma_{\pi}^2$ will be appropriately selected to adapt different scenarios. The smaller the variance is, the estimated gradient can be more accurate, but it would lead to a slower convergence rate, which would reflect in the mutual information fluctuation of the low SNR region.  However, when the variance is larger, the algorithm may converge to the local optimum, and the mutual information in the high SNR regime would have a significant loss. Therefore, in this paper, $\sigma_{\pi}$ is selected from $[0.01, 0.15]$.  

\mbox{Fig. \ref{MI}} shows the mutual information results of several precoder design methods with different channels and modulation orders, 
where $\mathbf{H}_1= \begin{bmatrix}
	2&1\\
	1&1
\end{bmatrix}$ and $\mathbf{H}_2=\begin{bmatrix}
1&0.5j&0.3\\
-0.5j&1.5&-0.1j\\
0.3&0.1j&0.5
\end{bmatrix}$. We regard the DNN-based precoders design scheme with the complete channel model proposed by \cite{jing} as the model-aware method, compared with the model-free method proposed in this paper. Analyzing \mbox{Fig. 3}, in the low SNR region, the mutual information maximization problem is approximated to the power allocation problem. With the SNR increasing, the power allocation method MWF has a large loss for giving up the part of searching space related to the right singular matrix of the precoder. In this case, the precoders design method based on AE shows a large gain and the model-free method can be almost consistent with the model-aware method since the estimated gradient in this approach is close to the true one as much as possible by adjusting $\sigma_{\pi}$.  It can also be seen that the performance of the model-free method with Sigmoid is slightly worse than that with Softmax, because it maximizes the lower bound of the mutual information rather than the mutual information itself.


\begin{figure}
	\centering{\includegraphics[width = 2.8in]{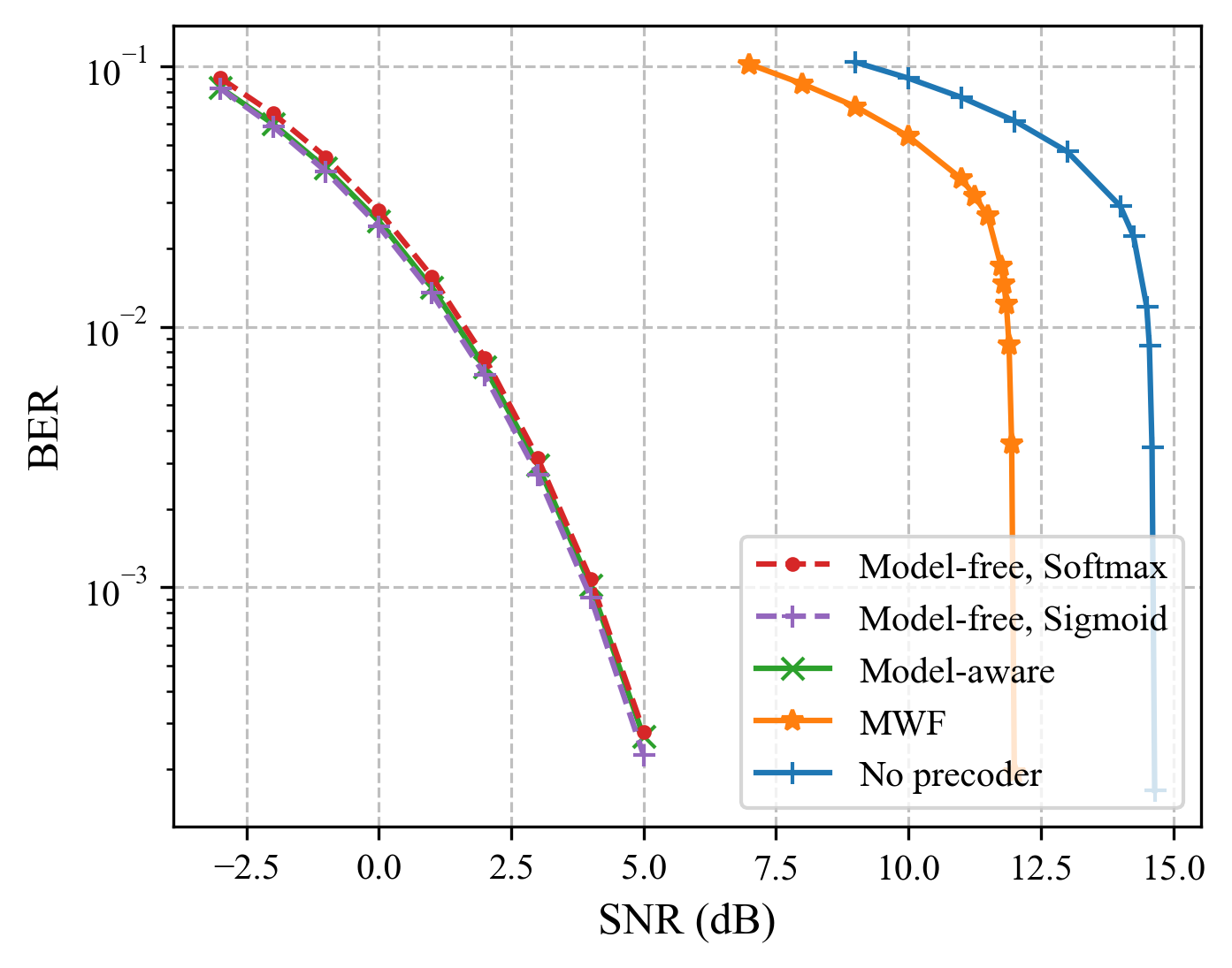}}
	\caption{BER with different precoding methods under $\mathbf{H}_1$, BPSK.}
	\label{ber}
\end{figure}

\mbox{Fig. \ref{ber}} shows the BER performance of the proposed method with $\mathbf{H}_1$ and BPSK modulation. In the simulation, the (648, 486) low-density parity-check (LDPC) code in the IEEE 802.11 is adopted for error correction and the maximum a posterior probability (MAP) criterion is used for detection. The iteration between the MAP detector and the LDPC decoder is 5. In this simulation, both the model-free and the model-aware methods are trained at SNR $=2$ dB. We observe that the performance of the model-free (no matter with Softmax or Sigmoid activation function) methods approach the BER of the model-aware method closely, both of which have a significant gain over the case of MWF or no precoder. This is consistent with the conclusion of the mutual information results.


\section{Conclusion}
In this paper, we solved the linear precoding design problem for finite alphabet inputs in MIMO systems without a channel model. Taking the advantages of the model-free network based AE, the alternating training on the receiver and the precoder was performed through the exact and estimated gradients, respectively. The proposed method obtained not only the precoders for maximizing the mutual information between channel inputs and outputs, but also the corresponding receiver, alleviating the high requirement of the existing methods on the perfect channel state information.  The simulation results showed that this no-channel-model design method achieved the performance of the complete-channel-model design method in terms of mutual information and bit error rate, offering practical insights for linear MIMO precoders design.

\bibliographystyle{IEEEtran}
\bibliography{IEEEabrv,myrefs}
\vspace{12pt}

\end{document}